\documentclass[aps,prb,reprint,groupedaddress,showpacs]{revtex4-1}
\usepackage{amsmath}
\usepackage{amssymb}
\usepackage{graphicx}
\usepackage{array}
\usepackage{dcolumn}
\usepackage{subfigure}
\usepackage{color}
\usepackage{float}
\usepackage{ulem}
\usepackage[hidelinks=true]{hyperref}
\hypersetup{
  colorlinks   = true, 
  urlcolor     = blue, 
  linkcolor    = blue, 
  citecolor   = red 
}

\usepackage[T1]{fontenc}
\usepackage[latin9]{inputenc}
\usepackage{multirow}

\makeatletter

\providecommand{\tabularnewline}{\\}

\makeatother

\usepackage{babel}
\begin{document}
\preprint{APS/123-QED}
\title{Effect of Spin-Orbit Coupling in non-centrosymmetric half-Heusler
alloys}
\author{Kunal Dutta} 
\author{Subhadeep Bandyopadhyay}
\thanks{current address: Theoretical Materials Physics, CESAM, University of Li\`ege, Li\`ege, Belgium }  
\author{Indra Dasgupta}
\email{sspid@iacs.res.in}
\affiliation{School of Physical Sciences, Indian Association for the Cultivation
of Science, 2A and 2B Raja S.C. Mullick Road, Jadavpur, Kolkata 700
032, India}
\date{\today}
\begin{abstract}
Spin-orbit coupled electronic structure of two representative non-polar
half-Heusler alloys, namely 18 electron compound CoZrBi and 8 electron
compound SiLiIn have been studied in detail. An excursion through
the Brillouin zone of { these} alloys from one high symmetry point to
the other revealed rich local symmetry of the associated wave vectors
resulting in non-trivial spin splitting of the bands and consequent
diverse spin textures in the presence of spin-orbit coupling. Our
first principles calculations supplemented with low energy $\boldsymbol{k.p}$
model Hamiltonian revealed the presence of linear Dresselhaus effect
at the X point having $D_{2d}$ symmetry and Rashba effect with both
linear and non-linear terms at the L point with $C_{3v}$ point group
symmetry. Interestingly we have also identified non-trivial Zeeman
spin splitting at the non-time reversal invariant W point and a pair
of non-degenerate bands along the path $\Gamma$ to L displaying vanishing
spin polarization due to the non-pseudo polar point group symmetry
of the wave vectors. { Further a comparative study of CoZrBi and SiLiIn suggest, in addition to the local symmetry of the wave vectors, important role of the participating orbitals in deciding the magnitude of the spin splitting of the bands}. Our calculations suggest half-Heusler compounds
with heavy elements displaying diverse spin textures may be ideal
candidate for spin valleytronics where spin textures can be controlled
by accessing different valleys around the high symmetry k-points. 
\end{abstract}
\maketitle
\section{\label{sec:level1-1}INTRODUCTION}
In non-centrosymmetric systems, the non-vanishing gradient of the
electrostatic potential results in a momentum dependent magnetic field
$\boldsymbol{\Omega\left(k\right)}$ in the rest frame of the electron.
The coupling of this field with the spin $\vec{\sigma}$ of the electron
lifts the spin-degeneracy of the bands in an otherwise non-magnetic
system. The resulting spin-orbit coupled Hamiltonian is given by,
$H_{SOC}=\boldsymbol{\Omega\left(k\right).\sigma}$. Depending on
the symmetry, $\boldsymbol{\Omega\left(k\right)}$ may have both linear
and higher order k-dependent terms. The momentum dependent field $\boldsymbol{\Omega\left(k\right)}$
that locks the electron's spin direction to its momentum not only
remove the spin degeneracy of the bands but also leads to complex
spin textures in the reciprocal space. Spin texture(ST) is the expectation
value of the spin operator $\left\langle \overrightarrow{S{_{n}}}\left(k\right)\right\rangle $
in a given bloch wave function $u{_{n}}\left(k\right)$ around a specific
k-point, where n is the band index. Spin textures depending on the
symmetry may display Rashba \citep{rashba1960properties}, Dresselhaus
\citep{Dresselhaus}, persistent \citep{PSH}, { radial} \citep{Radial} or more complex spin
configurations in the momentum space. In addition to the well studied
Dresselhaus and Rashba effect that leads to splitting of non-degenerate
bands with characteristic ST in non-centrosymmetric systems, there
are also other possibilities where spin degeneracy can be lifted in
non-magnetic, non-centrosymmetric systems due to the presence of $\boldsymbol{\Omega\left(k\right)}$.
\\
 \indent It is well known that in materials either with intrinsic
magnetic ordering or in the presence of a magnetic field, time reversal(TR)
symmetry is broken, leading to splitting of energy bands with opposite
spins, referred to as, Zeeman spin splitting \citep{Zeeman1}. In
nonmagnetic, non-cetrosymmetric compounds, a combination of non-time
reversal invariant k-point and lack of inversion symmetry in the presence
of spin-orbit coupling(SOC) also leads to spin splitting of bands
in the momentum space similar to the Zeeman effect \citep{Zeeman2}.
Further it has been recently shown the possibility of spin splitting
of bands in a non-magnetic, non-centrosymmetric crystals in the presence
of SOC, where the split bands does not show any net spin polarization
around certain high symmetry points of the Brillouin zone (BZ) \citep{BSVSP}.
Such systems with band splitting having vanishing spin polarisation
offers the possibility of tuning spin polarization either with the
application of an electric field or strain that may be important for
spintronics application. \\
 \indent Earlier research on Dresselhaus effect have primarily focused
on materials in which the bulk exhibits inversion asymmetry. It was
originally proposed for {non-polar} zinc-blende semiconductors, where
the splitting of the band is proportional to $k^{3}$ \citep{Dresselhaus}. On the other hand { the Rashba effect was proposed for non-centrosymmetric polar wurtzite structure displaying linear splitting of bands}\citep{Bihlmayer}. The search for Rashba effect  was {initially} confined to 2D
surfaces, interfaces. In view of the above, Rashba splitting was
observed on surfaces of heavy metals, such as Au $\left(111\right)$
\citep{Au(111)} and Bi $\left(111\right)$ \citep{Bi(111)}, at the
surface of oxides such as $\mathrm{SrTiO_{3}}$ $\left(001\right)$
\citep{SrTiO3} and $\mathrm{KTaO_{3}}$ $\left(001\right)$ \citep{KTaO3},
on the two dimensional materials \citep{2D_1,2D_2,2D_3} , and on
hetero-structure interfaces such as $\mathrm{InGaAs/InAlAs}$ \citep{InGaAs/InAlAs}
and $\mathrm{LaAlO_{3}/SrTiO_{3}}$ \citep{LaAlO3/SrTiO3}. However
recent studies on bulk polar materials show large Rashba spin splitting
for e.g. in $\mathrm{BiTeX\left(X=Cl,Br,I\right)}$ \citep{BiTeX_1,BiTeX_2,BiTeX_3}
and $\mathrm{GeTe}$ \citep{GeTe_1,GeTe_2}. Recently it has been
reported that both Dresselhaus and Rashba spin splitting is also observed
in bulk ferroelectric oxide perovskites such as $\mathrm{BiAlO_{3}}$
\citep{BaAlO3} , $\mathrm{HfO}_{2}$ \citep{HfO2} and $\mathrm{nitride}$
perovskite $\mathrm{LaWN_{3}}$ \citep{LaWN3}. Further recent studies
suggest that the Rashba effect occurs not only in bulk for polar and
ferroelectric crystal structures, but may also occur in non-polar
crystals with polar point group in the BZ \citep{PhysRevB.104.104408}.
The Zeeman effect on the other hand was observed in the two-dimensional
$\mathrm{WSe}_{2}$ and $\mathrm{MoS}_{2}$ \citep{Yuan2013,PhysRevLett.108.196802},
and a large Zeeman splitting was also observed in bulk $\mathrm{OsC}$,
$\mathrm{WN_{2}}$, $\mathrm{SnTe}$ \citep{MeraAcosta2019} and in
the non-polar $\mathrm{GaAs}$ \citep{GaAs}. Band splitting with
vanishing spin polarisation was suggested to be realised in bulk nonsymmorphic
GaAs and symmorphic 2D-SnTe \citep{BSVSP}. \\
 \indent While crystallographic point group symmetry(CPGS) was originally
attributed to be responsible for the nature of spin textures (Rashba
or Dresselhaus type), very recently it was observed that the point
group symmetry of the wave vector (little group) \citep{PhysRevB.104.104408}
and the symmetry of the orbitals involved \citep{PhysRevB.103.014105}
play a crucial role in determining the spin textures. As a consequence,
in the same compound diverse spin textures can be realised around
different high symmetry k-points of the BZ depending on its little
group. Similarly different compounds at a particular k-point despite
having same little group, depending on the orbital character of the
bands may display diverse spin textures \citep{PhysRevB.103.014105}.
\\
 \indent In this paper, we have considered two representative semiconducting
half-Heusler compounds with heavy elements having 18 and 8 valance
electrons respectively. The unique non-polar crystal structure of
the half-Heusler compounds where the ternary half-Heusler is a combination
of two binaries, one with centrosymmetric and the other with non-centrosymmetric
structure will be shown to host variety of spin textures depending
on the symmetry of the k-points in the BZ. Our calculations clearly
identify the importance of the little group of the chosen k-point in determining
the nature of the spin textures. A comparative study of the two half-Heusler
compounds show that the orbital character of the bands are important
in deciding the strength of the SOC induced band splitting. \\
 \indent The rest of the paper is organized as follows. Sec. \ref{sec:level2}
is devoted to the description of the structural properties of non-polar
half-Heusler compounds and details of the computational methods. In
Sec. \ref{sec:level3} we have discussed the results of our first
principles electronic structure calculations. Sec. \ref{sec:level4}
is devoted to the description of the nature of band splitting and
consequent spin textures at various high symmetry k-points of the
BZ. The results of our first principle calculations have been analysed
here in the framework of a low energy $\boldsymbol{k.p}$ model. Finally,
conclusions are presented in Sec. \ref{sec:level5}.
\begin{figure}[t]
\includegraphics[width=7.5cm,height=5.8cm,keepaspectratio]{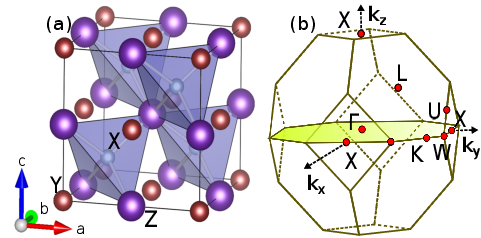}\caption{\label{fig:figure1} (a) Crystal structure and (b) BZ of bulk half-Heusler
compounds. The half-Heusler compound have XYZ composition, where X(Co,
Si) have higher valence compared to Y(Zr, Li) and, Y and Z(Bi, In)
form a rock salt structure.}
\end{figure}
\begin{table}[h]
\caption{\label{tab:table1}The lattice constant and atomic positions for bulk
half-Heusler compounds.}
\begin{ruledtabular}
\begin{tabular}{l>{\raggedright}p{2.25cm}ccccc}
Space group & Bulk half-Heusler & a$\left(\mathring{A}\right)$ & Site & x & y & z\tabularnewline
\hline 
\multirow{7}{*}{$\begin{array}{c}
F\bar{4}3m\\
\left(216\right)
\end{array}$} & \multirow{3}{2.25cm}{18 electron CoZrBi} & \multirow{3}{*}{6.23\citep{YAZDANIKACHOEI2020154287}} & Zr(4a) & 0.00 & 0.00 & 0.00\tabularnewline
 &  &  & Co(4c) & 0.25 & 0.25 & 0.25\tabularnewline
 &  &  & Bi(4b) & 0.50 & 0.50 & 0.50\tabularnewline
 & \multicolumn{6}{c}{}\tabularnewline
 & \multirow{3}{2.25cm}{8 electron SiLiIn} & \multirow{3}{*}{6.31\citep{Sahni2019}} & Li(4a) & 0.00 & 0.00 & 0.00\tabularnewline
 &  &  & Si(4c) & 0.25 & 0.25 & 0.25\tabularnewline
 &  &  & In(4b) & 0.50 & 0.50 & 0.50\tabularnewline
\end{tabular}
\end{ruledtabular}
\end{table}
\section{\label{sec:level2}STRUCTURAL AND COMPUTATIONAL DETAILS}
The half-Heusler compound XYZ crystallize in the face-centered cubic
structure with one formula unit per primitive unit cell, as shown
in Fig. \ref{fig:figure1}(a). In normal half-Heusler alloys, X and
Y are transition metals with X being higher valence element in comparison
to Y, and Z is a sp valent element. The space group is $F\bar{4}3m$
(No. 216). In the conventional stable structure Y and Z atoms are
located at $4a(0,0,0)$ and $4b(\frac{1}{2},\frac{1}{2},\frac{1}{2})$
positions forming the rock salt structure arrangement and the X atom
is located in the { tetrahedrally} coordinated pocket, at one of the cube
center positions $4c(\frac{1}{4},\frac{1}{4},\frac{1}{4})$ leaving
the other $4c(\frac{3}{4},\frac{3}{4},\frac{3}{4})$ empty, resulting
in the absence of inversion symmetry. We shall consider a representative
semiconducting 18 valence electron half-Heusler compound CoZrBi \citep{YAZDANIKACHOEI2020154287}.
In addition, we have also considered a representative semiconducting
sp valent compound SiLiIn \citep{Sahni2019} featuring half-Heusler
structure with 8 valence electrons. \\
 \indent The calculations presented in this paper have been carried
out using Vienna $ab$ initio simulation package (VASP) \citep{PhysRevB.47.558,PhysRevB.54.11169}within
density-functional theory (DFT) using the supplied projector augmented-wave
pseudo potentials \citep{PhysRevB.50.17953,PhysRevB.59.1758} and
Perdew-Burke-Ernzerhof generalized gradient approximation (GGA) \citep{PhysRevLett.77.3865}.
Here the energy cutoff has been set to 600 $\mathrm{eV}$ for the
calculations and a 10$\times$ 10 $\times$ 10 k-point mesh is used
for the self-consistent calculations using the Monkhorst grid for
k-point sampling. All the calculations are done with experimental
lattice constant. The details of the structure, including structural
parameters are summarised in Table \ref{tab:table1}.
\\
\begin{figure*}[!t]
\includegraphics[width=17.5cm,height=7.15cm,keepaspectratio]{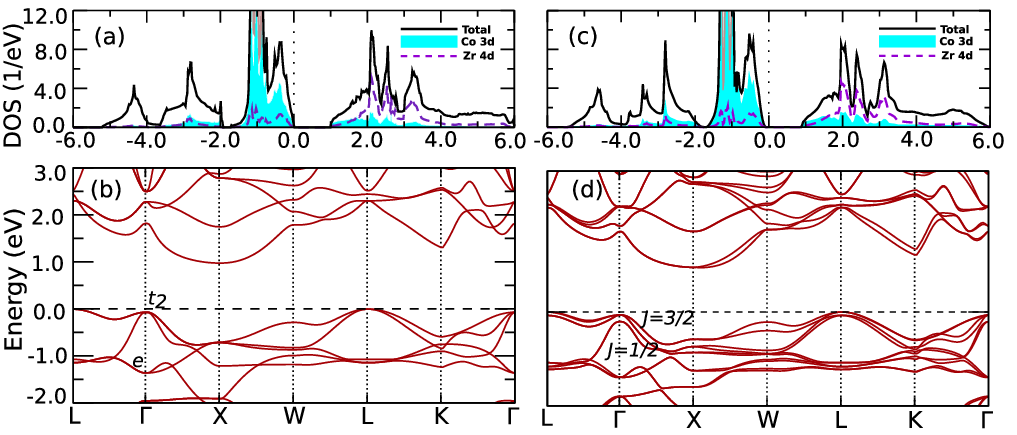}\caption{\label{fig:figure2}The density of states and electronic band structure
of the 18 electron half-Heusler compound CoZrBi, in the absence and
presence of SOC. In (a) and (c) the total and projected DOS for the
Co-3d and Zr-4d are shown without and with SOC respectively. (b) and
(d) Displays the band structure along various high symmetry points
of the BZ in absence and presence of SOC respectively. The band structures
of CoZrBi are plotted along the high-symmetry points L$(\frac{\pi}{a},\frac{\pi}{a},\frac{\pi}{a})$-$\Gamma{\left(0,0,0\right)}$-X$(0,\frac{2\pi}{a},0)$-W$(\frac{\pi}{a},\frac{2\pi}{a},0)$-L$(\frac{\pi}{a},\frac{\pi}{a},\frac{\pi}{a})$-K$(\frac{3\pi}{2a},\frac{3\pi}{2a},0)$-$\Gamma(0,0,0)$
of the BZ. The Fermi level is aligned to the valence-band maximum
with zero value in the energy axis.}
\end{figure*}
\indent Fig. \ref{fig:figure1}(b) represents the BZ of the face-centered
cubic half-Heusler compound. The various high symmetry points of the
BZ are $\Gamma$ at the center of the BZ, L at the center of each
hexagonal face, X at the center of each square face, W at each corner
formed from one square and two hexagons, K at middle of an edge joining
two hexagonal faces. Further X point has 6 fold-degeneracy, L point
has 8 fold degeneracy, W and K points have 12 fold degeneracy. In
the BZ, except W and K points all are time reversal(TR) invariant. A time reversal invariant k-point \citep{SzVajna2012} satisfies the condition $-\boldsymbol{k}+\boldsymbol{G}=\boldsymbol{k}$, where $\boldsymbol{G}$ is the reciprocal lattice vector.
\section{\label{sec:level3}ELECTRONIC STRUCTURE CALCULATIONS}
To begin with we have analysed the density of states without SOC for
the 18 electron compound CoZrBi obtained from DFT calculations. The
non-spin-polarized total as well as projected density of states of
Co-3d and Zr-4d have been shown in Fig. \ref{fig:figure2}(a). The
characteristic feature of the total DOS is a pair of bonding and antibonding
states resulting from the covalent hybridization of the higher valence
Co-d and lower valence Zr-d states separated by a 0.98 eV gap at the
Fermi level \citep{BRKNanda_2003}. Below the bonding states are the
Bi-p states separated by a p-d gap. Below the Bi-p state lies the
Bi-s state. The semiconducting nature of the compound can be understood
from the electron filling of the system. As the total number of valence
electrons of the system is 18, these are accommodated in the available
Bi-s, Bi-p and the bonding partner of the Co-d - Zr-d hybridised states.
The Co-d and Zr-d projected DOS reveal that the bonding states have
primary contribution from the 3d states of Co, while the antibonding
states are primarily composed of Zr-d states. \\
 \indent In Fig. \ref{fig:figure2}(b) we have shown the band structure
of CoZrBi around the Fermi level. An indirect band gap of 0.98 eV
is observed between the L point of the valence band and the X point
of the conduction band, in agreement with a previous report \citep{Zhu2018}.
Further due to the tetrahedral network, at the $\Gamma$ point, the
top of the valence band consisting of Co-Zr d-states split into 3-fold
degenerate $t_{2}$ and 2-fold degenerate $e$ states, with the latter
lying lower in energy. \\
 \indent Fig. \ref{fig:figure2}{[}(c) and (d){]} display DOS and
band structure of CoZrBi including SOC. The gross feature of the DOS
is very similar to that obtained without SOC, except for the presence
of additional splittings and the value of the d-d gap is now 0.96
eV. However the non trivial effect of the absence of inversion symmetry
upon inclusion of SOC is revealed in the band structure shown in Fig.
\ref{fig:figure2}(d). The effect of the SOC depends on the symmetry
of the paths in the reciprocal space. The spin degeneracy of the bands
along various high symmetry direction of the BZ are lifted. Of particular
interest is the top of the valence band at the $\Gamma$ point, where
SOC further splits the $t_{2}$ states into four fold degenerate $J=\frac{3}{2}$
and a two fold degenerate $J=\frac{1}{2}$ states with the latter
lying lower in energy in the tetrahedral environment. { Unlike the $t_2$ states, the $e$-states retain their degeneracy at the $\Gamma$ point.} 
\begin{figure*}[t]
\includegraphics[width=17.5cm,height=7.15cm,keepaspectratio]{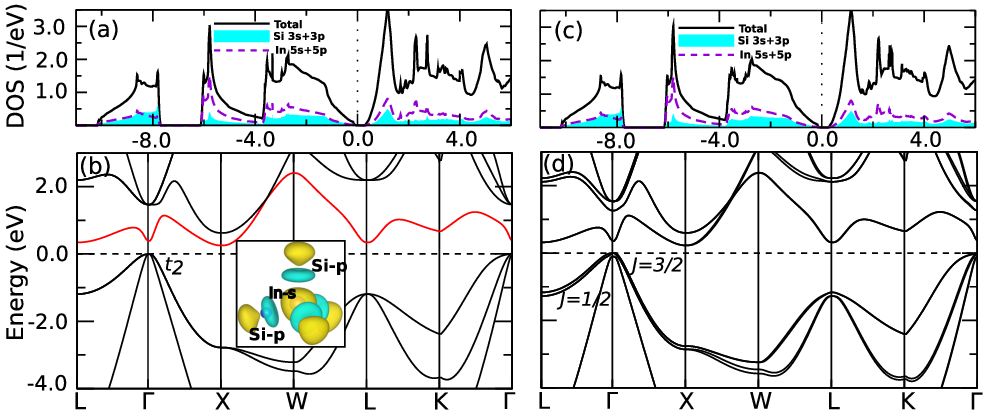}\caption{\label{fig:figure3}The density of states and electronic band structure
of the 8 electron half-Heusler compound SiLiIn, in the absence and
presence of SOC. In (a) and (c) the total and projected DOS for the
Si-3s+3p and In-5s+5p are shown without and with SOC respectively.
(b) and (d) Displays the band structure along various high symmetry
points of the BZ in absence and presence of SOC respectively. The
band structures of SiLiIn are plotted along the high-symmetry points
L$(\frac{\pi}{a},\frac{\pi}{a},\frac{\pi}{a})$-$\Gamma\left(0,0,0\right)$-X$(0,\frac{2\pi}{a},0)$-W$(\frac{\pi}{a},\frac{2\pi}{a},0)$-L$(\frac{\pi}{a},\frac{\pi}{a},\frac{\pi}{a})$-K$(\frac{3\pi}{2a},\frac{3\pi}{2a},0)$-$\Gamma(0,0,0)$
of the BZ. Fermi level is aligned to the valence-band maximum with
zero value in the energy axis. The inset in (b) represents the Wannier
function of the lowest conduction band.}
\end{figure*}
\indent The total as well as the projected DOS and the band structure
around the Fermi level without SOC for the 8 electron half-Heusler
SiLiIn is shown in Fig. \ref{fig:figure3}(a) and Fig. \ref{fig:figure3}(b)
respectively. Similar to the 18 electron compound the characteristic
feature of the DOS is a pair of bonding and antibonding states derived
from In s+p and Si s+p separated by a semiconducting gap of magnitude
0.12 eV. Below the bonding state lies the Si-s state. As the total
number of valence electrons of the system is 8, these are accommodated
in the Si-s state and bonding partner of the p-states below the Fermi
level. The plot of the band structure reveal an indirect band gap
of 0.12 eV between the $\Gamma$ point of the valence band and the
X point of the conduction band. The top of the valence band at the
$\Gamma$ point is the 3-fold degenerate p-band. The lowest conduction
band is predominantly sp$^{3}$ hybridized In-s - Si-p band (see Fig.
\ref{fig:figure3}(b) inset). \\
 \indent Finally the DOS and band structure of SiLiIn including SOC
is shown in Fig. \ref{fig:figure3}{[}(c),(d){]}. The gross feature
of the DOS is very similar to that without SOC shown in Fig. \ref{fig:figure3}(a)
except for additional splittings. As expected, around the $\Gamma$
point of the valence band, the three fold degenerate p-bands splits
into $J=\frac{3}{2}$ and $J=\frac{1}{2}$ states, where the splitting
is much smaller in comparison to the 18 electron compound.
\section{\label{sec:level4}Nature of band splitting and Spin Textures}
Next we have analysed the nature of the SOC induced splitting of the
bands near various high symmetry points of the BZ, in order to elucidate
the importance of local symmetry. { In the following, we have discussed in detail the spin orbital locked split bands and the novel spin textures displayed by them in the k-space at and around the TR invariant non-polar(X), polar(L) and non-time reversal invariant W points, which based on local symmetry are expected to display Dresselhaus effect, Rashba effect and Zeeman effect respectively.}
\subsection{Dresselhaus and Rashba effect}
\subsubsection{X point$\left(0,0,\frac{2\pi}{a}\right)$ }
\begin{figure*}
\includegraphics[width=17cm,height=8.3cm,keepaspectratio]{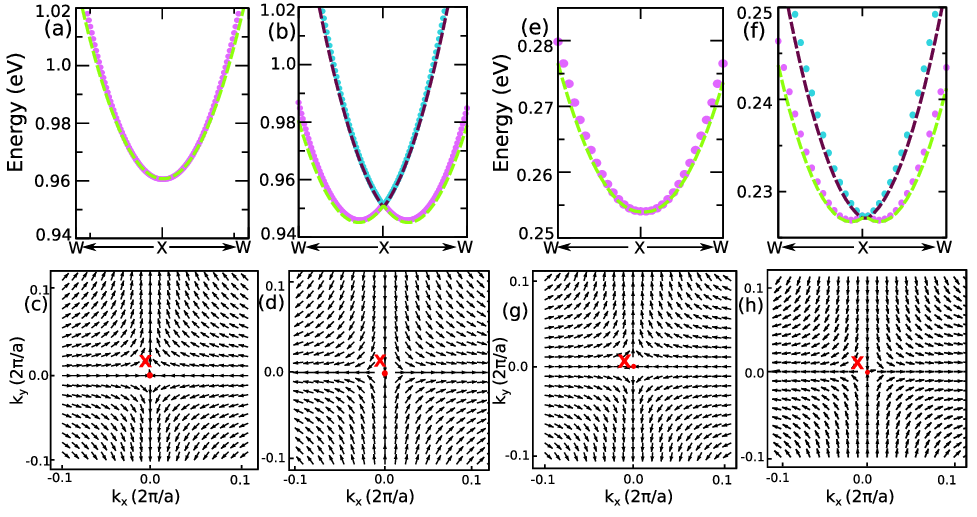}\caption{\label{fig:figure4} The band structure and ST of CoZrBi and SiLiIn
plotted in the $k_{z}=\frac{2\pi}{a}$ plane in a very narrow k-range
around the X point in the presence and absence of SOC. In (a) and
(b), the band structures of CoZrBi without and with SOC plotted along
$\frac{2\pi}{a}(0.3,0.0,1.0)$-$\frac{2\pi}{a}(0.0,0.0,1.0)$-$\frac{2\pi}{a}(0.0,0.3,1.0)$
path, which is along the path W$\leftarrow$X$\rightarrow$W. In (e)
and (f) the band structures of SiLiIn without and with SOC are plotted
along $\frac{2\pi}{a}(0.075,0.0,1.0)$-$\frac{2\pi}{a}(0.0,0.0,1.0)$-$\frac{2\pi}{a}(0.0,0.075,1.0)$
direction. The band structure obtained form the DFT calculation as
plotted with dots while the band structure obtained from the $\boldsymbol{k.p}$
model Hamiltonian are plotted with dashed lines. In (c) and (d), the
inner and outer branches of ST of CoZrBi around the X point in the
conduction band minimum(CBM) are shown. (g) and (h), displays the {same} spin-texture for SiLiIn.}
\end{figure*}

To begin with we have analysed the conduction band minimum(CBM) of
CoZrBi around the neighborhood of the high symmetry {non-polar} X point along
the path W$\left(\frac{\pi}{a},0,\frac{2\pi}{a}\right)$-X$\left(0,0,\frac{2\pi}{a}\right)$-W$\left(0,\frac{\pi}{a},\frac{2\pi}{a}\right)$
in the $k_{z}=\frac{2\pi}{a}$ plane. The DFT band structure for CoZrBi
in a narrow k-range along the above mentioned path, without and including
SOC are displayed in Fig. \ref{fig:figure4}(a) and \ref{fig:figure4}(b)
respectively. The band structure including SOC shows that the band
minimum is shifted from the X point in both directions, reminiscent
of Rashba-Dresselhaus effect. In order to identify the nature of the
spin splitting, the ST of the inner and outer branches of the CBM
around the X point has been shown in Fig. \ref{fig:figure4}(c) and
\ref{fig:figure4}(d) respectively. As shown in Fig. \ref{fig:figure4}{[}(c),
(d){]}, the angle between the $\boldsymbol{{k}}$ and the expectation
values of spin $\left\langle S_{x}\right\rangle $ and $\left\langle S_{y}\right\rangle $
varies with the direction and the $\langle S{}_{z}\rangle$ component
is absent. Along the $k_{x}$ and $k_{y}$ axis, the spin is parallel
to $\boldsymbol{{k}}$, while it is perpendicular to $\boldsymbol{{k}}$
along the diagonals. As expected the direction of the spin textures
are opposite for the inner and outer branches as shown in \ref{fig:figure4}(c)
and \ref{fig:figure4}(d) respectively. These are characteristic signatures
of linear Dresselhaus effect. In order to understand our DFT results
next we have derived a low energy $\boldsymbol{k.p}$ model Hamiltonian.
\begin{table}[b]
\caption{\label{tab:table2}Symmetry operations of $D_{2d}$ point group.}
\begin{ruledtabular}
\begin{tabular}{lcc}
 & X point & \multicolumn{1}{c}{}\tabularnewline
\hline 
$\begin{array}{l}
\text{Symmetry}\\
\text{operation}
\end{array}$ & $\{k_{x},k_{y},k_{z}\}$ & $\{\sigma_{x},\sigma_{y},\sigma_{z}\}$\tabularnewline
\hline 
$C_{2}\left(z\right)=-i\sigma{}_{z}$ & $\left\{ -k_{x},-k_{y},k_{z}\right\} $ & $\left\{ -\sigma_{x},-\sigma_{y},\sigma_{z}\right\} $\tabularnewline
$C_{2}^{\prime}\left(x\right)=-i\sigma{}_{x}$ & $\left\{ k_{x},-k_{y},k_{z}\right\} $ & $\left\{ \sigma_{x},-\sigma_{y},-\sigma_{z}\right\} $\tabularnewline
$C_{2}^{\prime}\left(y\right)=-i\sigma{}_{y}$ & $\left\{ -k_{x},k_{y},k_{z}\right\} $ & $\left\{ -\sigma_{x},\sigma_{y},-\sigma_{z}\right\} $\tabularnewline
$S_{4}(z)=e{}^{i\frac{\pi}{4}\sigma{}_{z}}$ & $\left\{ k_{y},-k_{x},k_{z}\right\} $ & $\left\{ -\sigma_{y},\sigma_{x},\sigma_{z}\right\} $\tabularnewline
$M_{d1}=-i\left(\frac{-\sigma{}_{x}+\sigma{}_{y}}{\sqrt{2}}\right)$ & $\left\{ k_{y},k_{x},k_{z}\right\} $ & $\left\{ -\sigma_{y},-\sigma_{x},-\sigma_{z}\right\} $\tabularnewline
$M_{d2}=-i\left(\frac{\sigma{}_{x}+\sigma{}_{y}}{\sqrt{2}}\right)$ & $\left\{ -k_{y},-k_{x},k_{z}\right\} $ & $\left\{ \sigma_{y},\sigma_{x},-\sigma_{z}\right\} $\tabularnewline
\end{tabular}
\end{ruledtabular}
\end{table}
\indent The point-group, symmetry around the X$(0,0,\frac{2\pi}{a})$
point is $D_{2d}$ having twofold $C_{2}\left(z\right)$ rotation
around the z axis(principle axis), twofold rotation perpendicular
to the principle axis$\left(C_{2}^{\prime}\left(x\right),C_{2}^{\prime}\left(y\right)\right)$,
reflection in the dihedral plane$\left(M_{d1},M_{d2}\right)$, four
fold rotation followed by reflection through a plane perpendicular
to the principle axis $S_{4}\left(z\right)$. All the symmetries are
listed in Table \ref{tab:table2}. The symmetry operations listed
in Table \ref{tab:table2}, keep the linear Dresselhaus Hamiltonian
$H_{D}\left(k\right)$ invariant. Further Table \ref{tab:table2}
reveal that out of plane spin component is zero, as no linear combination
of $k_{x}$ and $k_{y}$ with $\sigma_{z}$ is invariant under the
symmetry operations. The effective Hamiltonian is, 
\begin{equation}
H_{X}^{}(k)=H_{0}(k)+\alpha_{D}(\sigma_{x}k_{x}-\sigma_{y}k_{y})\label{eq:equation1}
\end{equation}
\indent where $H_{0}$ is the Hamiltonian of the free electrons with
the dispersion $E_{0}\left(k\right)=\frac{\hbar^{2}}{2m^{*}}\left(k_{x}^{2}+k_{y}^{2}+\left(\frac{2\pi}{a}\right)^{2}\right)$
and $\alpha_{D}$ is the Dresselhaus coupling constant. Diagonalisation
of Eqn. \ref{eq:equation1} yield, $E(k)^{\pm}=E_{0}\pm\alpha_{D}\sqrt{\left(k_{x}^{2}+k_{y}^{2}\right)}$.
The band structure obtained from the model Hamiltonian calculation
around X point without and including SOC are shown with dashed lines
in Figs. \ref{fig:figure4}(a) and \ref{fig:figure4}(b), respectively
and it agrees well with the DFT calculations. The values of the Dresselhaus
parameter $\left(\alpha{}_{D}\right)$ obtained from the model Hamiltonian
is $\alpha_{D}$ = 0.26 $\mathrm{eV}\mathring{A}$. The value of $\alpha{}_{D}$
calculated as twice the ratio between the shift in energy and momentum
from DFT calculation, $\alpha_{D}=\frac{2\delta E}{\delta k}$=0.25
eV$\mathring{A}$, is in a good agreement with that obtained from
the $\boldsymbol{k.p}$ calculations. \\
\begin{figure*}[t]
\includegraphics[width=17cm,height=10.2cm,keepaspectratio]{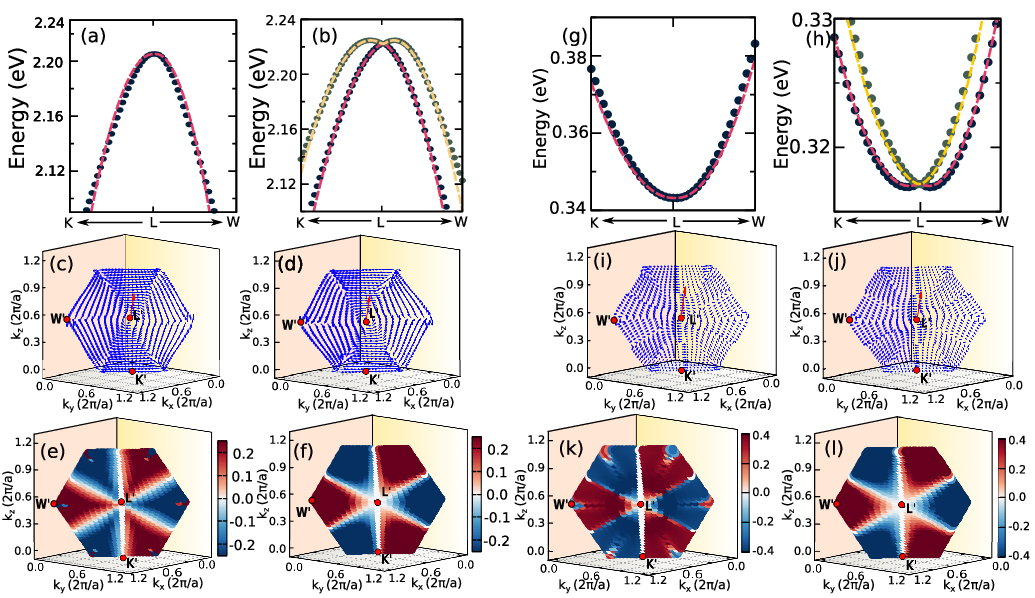}\caption{\label{fig:figure5} ((a), (b)) Band structure without and with SOC {for the conduction band}
for CoZrBi is plotted along $\frac{2\pi}{a}(0.563,0.563,0.374)$-$(\frac{2\pi}{a}(0.5,0.5,0.5)$-$\frac{2\pi}{a}(0.623,0.377,0.5)$
which lies along the path K$\leftarrow$L$\rightarrow$W. Band structure
obtained form DFT is plotted with dots and the band structure obtained
from the $\boldsymbol{k.p}$ model Hamiltonian is plotted with dashed
lines. ((c), (d)) ST of inner and outer branches around L$^{\prime}$
point for CoZrBi for the conduction band obtain from DFT calculation.
((e), (f)) Out-of-plane spin component of CoZrBi in the plane defined
by $k_{x}+k_{y}+k_{z}=\frac{3\pi}{a}$, obtained from DFT and model
Hamiltonian respectively. ((g), (h)) Band structure without and with
SOC {for the conduction band} for SiLiIn  is plotted along $\frac{2\pi}{a}(0.531,0.531,0.437)$-$(\frac{2\pi}{a}(0.5,0.5,0.5)$-$\frac{2\pi}{a}(0.56,  0.44 ,0.5)$
which lies along the path K$\leftarrow$L$\rightarrow$W. ((i),(j))
ST of inner and outer branches around L$^{\prime}$ point for SiLiIn
in the conduction band. (k), (l) Out-of-plane spin component of SiLiIn
in the plane defined by $k_{x}+k_{y}+k_{z}=\frac{3\pi}{a}$, obtained
from DFT and model Hamiltonian respectively.}
\end{figure*}
\indent Similarly, for the 8 electron system SiLiIn, the band structure
of the CBM around the X point, in the absence and presence of SOC
is displayed in Fig. \ref{fig:figure4}{[}(e), (f){]} respectively.
The band splitting seen in Fig. \ref{fig:figure4}(f) suggests Rashba-Dresselhaus
effect. A comparison of ST shown in Fig. \ref{fig:figure4}{[}(c),
(d){]} with Fig. \ref{fig:figure4}{[}(g), (h){]} reveal that the
ST for 18 and 8 electrons are identical. As expected from the nature of band splitting the Dresselhaus parameter $\alpha{}_{D}$  is small and estimated to
be 0.08 $\mathrm{eV}\mathring{A}$. { This small splitting is attributed to the participation of the sp$^{3}$ hybridized states for SiLiIn in contrast to the d states for CoZrBi, where the strength of SOC is expected to be higher for the latter.}
\newline
\indent Using the model Hamiltonian Eqn. \ref{eq:equation1}, we have calculated the band structure
without and with SOC and is shown with dotted line in Fig. \ref{fig:figure4}{[}(e),
(f){]}, which agrees well with the band structure obtained from the
DFT calculations. Calculated Dresselhaus parameter from the model
Hamiltonian, $\alpha_{D}=0.08$ eV$\mathring{A}$, is in excellent
agreement with the DFT estimate.
\begin{table*}[!t]
\caption{\label{tab:table3} Symmetry operations of $C_{3v}$ point group.}
\begin{ruledtabular}
\begin{tabular}{lcc}
\multicolumn{3}{c}{L point}\tabularnewline
\hline 
Symmetry operation & $\left\{ k_{x}^{\prime},k_{y}^{\prime},k_{z}^{\prime}\right\} $ & $\left\{ \sigma_{x}^{\prime},\sigma_{y}^{\prime},\sigma_{z}^{\prime}\right\} $\tabularnewline
\hline 
$C_{3}^{+}\left(\left[111\right]\right)=e^{-i\frac{\pi}{3}\sigma_{z}^{\prime}}$ & $\left\{ \begin{array}{c}
\left(-\frac{1}{2}k_{x}^{\prime}+\frac{\sqrt{3}}{2}k_{y}^{\prime}\right),\left(-\frac{\sqrt{3}}{2}k_{x}^{\prime}-\frac{1}{2}k_{y}^{\prime}\right),k_{z}^{\prime}\end{array}\right\} $ & $\left\{ \begin{array}{c}
\left(-\frac{1}{2}\sigma_{x}^{\prime}+\frac{\sqrt{3}}{2}\sigma_{y}^{\prime}\right),\left(-\frac{\sqrt{3}}{2}\sigma_{x}^{\prime}-\frac{1}{2}\sigma_{y}^{\prime}\right),\sigma_{z}^{\prime}\end{array}\right\} $\tabularnewline
$C_{3}^{-}\left(\left[111\right]\right)=e^{i\frac{\pi}{3}\sigma_{z}^{\prime}}$ & $\left\{ \begin{array}{c}
\left(-\frac{1}{2}k_{x}^{\prime}-\frac{\sqrt{3}}{2}k_{y}^{\prime}\right),\left(\frac{\sqrt{3}}{2}k_{x}^{\prime}-\frac{1}{2}k_{y}^{\prime}\right)\end{array},k_{z}^{\prime}\right\} $ & $\left\{ \begin{array}{c}
\left(-\frac{1}{2}\sigma_{x}^{\prime}-\frac{\sqrt{3}}{2}\sigma_{y}^{\prime}\right),\left(\frac{\sqrt{3}}{2}\sigma_{x}^{\prime}-\frac{1}{2}\sigma_{y}^{\prime}\right)\end{array},\sigma_{z}^{\prime}\right\} $\tabularnewline
$\sigma=-i\left(-\frac{1}{2}\sigma_{x}^{\prime}+\frac{\sqrt{3}}{2}\sigma_{y}^{\prime}\right)$ & $\left\{ \begin{array}{c}
\left(\frac{1}{2}k_{x}^{\prime}+\frac{\sqrt{3}}{2}k_{y}^{\prime}\right),\left(\frac{\sqrt{3}}{2}k_{x}^{\prime}-\frac{1}{2}k_{y}^{\prime}\right),k_{z}^{\prime}\end{array}\right\} $ & $\left\{ \begin{array}{c}
\left(-\frac{1}{2}\sigma_{x}^{\prime}-\frac{\sqrt{3}}{2}\sigma_{y}^{\prime}\right),\left(-\frac{\sqrt{3}}{2}\sigma_{x}^{\prime}+\frac{1}{2}\sigma_{y}^{\prime}\right),-\sigma_{z}^{\prime}\end{array}\right\} $\tabularnewline
$\sigma^{\prime}=-i\left(-\frac{1}{2}\sigma_{x}^{\prime}-\frac{\sqrt{3}}{2}\sigma_{y}\prime\right)$ & $\left\{ \begin{array}{c}
\left(\frac{1}{2}k_{x}^{\prime}-\frac{\sqrt{3}}{2}k_{y}^{\prime}\right),\left(-\frac{\sqrt{3}}{2}k_{x}^{\prime}-\frac{1}{2}k_{y}^{\prime}\right),k_{z}^{\prime}\end{array}\right\} $ & $\left\{ \begin{array}{c}
\left(-\frac{1}{2}\sigma_{x}^{\prime}+\frac{\sqrt{3}}{2}\sigma_{y}^{\prime}\right),\left(\frac{\sqrt{3}}{2}\sigma_{x}^{\prime}+\frac{1}{2}\sigma_{y}^{\prime}\right),-\sigma_{z}^{\prime}\end{array}\right\} $\tabularnewline
$\sigma^{\prime\prime}=-i\sigma_{x}^{\prime}$ & $\left\{ -k_{x}^{\prime},k_{y}^{\prime},k_{z}^{\prime}\right\} $ & $\left\{ \sigma_{x}^{\prime},-\sigma_{y}^{\prime},-\sigma_{z}^{\prime}\right\} $\tabularnewline
\end{tabular}
\end{ruledtabular}
\end{table*}
\subsubsection{L point$\left(\frac{\pi}{a},\frac{\pi}{a},\frac{\pi}{a}\right)$ }
Next we have focused around the {polar} L point of the conduction band of the 18 electron half-Heusler compound CoZrBi that features a local maximum. The DFT band structure
without and with SOC is plotted in a narrow $k$-range around the
neighbourhood of the high symmetry L$\left(\frac{\pi}{a},\frac{\pi}{a},\frac{\pi}{a}\right)$
point along W$\left(\frac{2\pi}{a},0,\frac{\pi}{a}\right)$ and K$\left(\frac{3\pi}{2a},\frac{3\pi}{2a},0\right)$
directions as shown in Fig. \ref{fig:figure5}(a) and \ref{fig:figure5}(b)
respectively. The nature of the SOC induced band splitting into two
branches (see Fig. \ref{fig:figure5}(b)) indicates the presence of
either Rashba or Dresselhaus effect. We shall confirm the nature of
the band splitting by calculating the spin textures in the framework
of DFT supplemented with symmetry analysis within the $\boldsymbol{k.p}$
model. In order to facilitate the plot of ST we have considered a
plane $\perp$ to the $\left(111\right)$ direction, in such a way
so that L point is at the origin of this plane given by $k_{x}+k{}_{y}+k{}_{z}=\frac{3\pi}{a}$.
We define a co-ordinate system such that $k_{x}^{\prime}$ and $k_{y}^{\prime}$
are lying in the plane while $k_{z}^{\prime}$ is along $\left(111\right)$
direction. The new coordinate system $k_{x}^{\prime}$, $k_{y}^{\prime}$,
and $k_{z}^{\prime}$ is related to $k_{x}$, $k_{y}$, and $k_{z}$
by a shift of origin and rotation preserving the local point group
symmetry. The resulting unit vectors are $\widehat{k_{x}^{\prime}}=\frac{\widehat{k_{x}}}{\sqrt{2}}-\frac{\widehat{k_{y}}}{\sqrt{2}}$
, $\widehat{k_{y}^{\prime}}=\frac{\widehat{k_{x}}}{\sqrt{6}}+\frac{\widehat{k_{y}}}{\sqrt{6}}-\frac{2\widehat{k_{z}}}{\sqrt{6}}$
and $\widehat{k_{z}^{\prime}}=\frac{1}{\sqrt{3}}\widehat{k_{x}}+\frac{1}{\sqrt{3}}\widehat{k_{y}}+\frac{1}{\sqrt{3}}\widehat{k_{z}}$.
As a consequence, the co-ordinate of the high symmetry points in this
plane are {L$^{\prime}$}$\left(0,0,0\right)$, {W$^{\prime}$}$\left(\frac{\sqrt{2}\pi}{a},0,0\right)$
and {K$^{\prime}$}$\left(0,\frac{\sqrt{6}\pi}{2a},0\right)$. The
corresponding spin textures are calculated for the two branches in
the above mentioned plane around the L$^{\prime}$ point are shown
in Fig. \ref{fig:figure5}(c) and \ref{fig:figure5}(d) respectively.
The in-plane spin components exhibit distinct chiral configuration
as expected for the Rashba ST, while the presence of the out-of-plane
spin components with distinct pattern {[}Fig. \ref{fig:figure5}(e)
and (f){]} suggests that higher order k terms may be involved. Moving
from the inner to the outer branch direction, the chirality changes
from clockwise to counter-clockwise. In both the inner and outer branches,
the spin is orthogonal to the wave vector $\vec{k}$, which is typical
of Rashba-type SOC. The Rashba parameter estimated from the DFT calculation
is found to be $\alpha_{R}$=0.37 eV$\mathring{A}$. \\
\indent The $\boldsymbol{k.p}$ model Hamiltonian is constructed
by preserving $C_{3v}$ symmetry at the L point. The $\boldsymbol{k.p}$
Hamiltonian is defined in the reciprocal space ($k_{x}^{\prime}$,
$k_{y}^{\prime}$, $k_{z}^{\prime}$) as mentioned above, where the
three-fold rotation ($C_{3}$) occurs around the trigonal axis $k_{z}^{\prime}$
(parallel to the {[}111{]} direction) in both clockwise ($C_{3}^{+}$)
and anticlockwise ($C_{3}^{-}$) directions. The point group symmetry
also includes three mirror planes. One mirror plane lies in the $k_{y}^{\prime}$-$k_{z}^{\prime}$
plane and is defined as $\sigma^{\prime\prime}$, while the other
two mirror planes can be obtained by applying $C_{3}$ and $C_{3}^{2}$
operations to the initial mirror plane, and they are defined as $\sigma$
and $\sigma^{\prime}$, respectively. Under these symmetry operations,
the momentum and spin operators undergo transformation as listed in
Table \ref{tab:table3}. Thus, the symmetry-adapted model Hamiltonian
for the conduction band at the L point is,
\begin{eqnarray}
H_{L}^{C}\left({k}\right) & = & H_{0}\left(k\right)+H_{SOC}(k)\nonumber \\
 & = & -\frac{\hbar^{2}}{2m^{*}}\left(k_{x}^{\prime2}+k_{y}^{\prime2}\right)+\alpha_{R}\left(\sigma_{y}^{\prime}k_{x}^{\prime}-\sigma_{x}^{\prime}k_{y}^{\prime}\right)\nonumber \\
 &  & +\gamma\left(k_{x}^{\prime3}-3k_{x}^{\prime}k_{y}^{\prime2}\right)\sigma_{z}^{\prime}\label{eq:equation2}
\end{eqnarray}
The cubic term in the effective Hamiltonian is included to explain
the out-of-plane component $\sigma_{z}^{\prime}$ in the spin texture.
The band structure without and with SOC obtained from the model Hamiltonian
is shown with dotted line in Fig. \ref{fig:figure5}(a) and Fig. \ref{fig:figure5}(b)
respectively, is in good agreement with the DFT band structure. The
fitted Rashba parameter $\alpha_{R}$=0.38 eV$\mathring{A}$ and $\gamma=-0.32$
$\mathrm{\mathrm{eV}}\mathring{A^{3}}$ agrees well with the DFT estimate.
Further the out of the plane component of ST obtained from DFT calculations
is in reasonable agreement with that obtained from the model Hamiltonian
(see Fig. \ref{fig:figure4}(e) and \ref{fig:figure4}(f)) suggesting
the robustness of the low-energy $\boldsymbol{k.p}$ model Hamiltonian{[}Eqn.
\ref{eq:equation2}{]}. \\
\indent The band structure around the neighbourhood of the L point
without and with SOC for the s-p half-Heusler SiLiIn is presented
in Fig. \ref{fig:figure5}(g) and (h) respectively. The band structure
of the 8 electron compound has a minimum at the L-point in contrast
to a local maximum for the 18 electron compound due to the involvement of
sp$^{3}$ hybridized state and the calculated Rashba parameter, $\alpha_{R}$
= 0.14 eV$\mathring{A}$, which is much smaller compared to the 18
electron compound. The spin textures of the inner and outer branches
around the L$^{\prime}$ point of the lowest conduction band is shown
in Fig. \ref{fig:figure5} (i) and (j). In-plane spin components have
a pronounced chiral spin configuration, whereas the presence of the
out-of-plane spin component indicate the presence of higher-order
k terms as discussed before for the 18 electron compound{[}see Fig.
\ref{fig:figure5}(k) and (l){]}. \\
 \indent Since the point group symmetry of SiLiIn is identical to
18 electron half-Heusler compound CoZrBi, we can easily obtain the
model Hamiltonian for the 8 electron system around the L$^{\prime}$
point, where  $H_{0}{\left(k\right)}=\frac{\hbar^{2}}{2m^{*}}\left(k_{x}^{\prime2}+k_{y}^{\prime2}\right)$ and the $H_{SOC}$ is identical to Eqn.   \ref{eq:equation2}.The band structure obtained from the model Hamiltonian agree well with the DFT band dispersion shown with dotted lines in Fig. \ref{fig:figure5}(
(g), (h)). The fitted parameters are $\alpha_{R}$ = 0.15 $\mathrm{\mathrm{eV}}\mathring{A}$
and $\gamma=-0.51$ $\mathrm{\mathrm{eV}}\mathring{A^{3}}$. The obtained
Rashba parameter $\alpha_{R}$ agrees well with the DFT estimate.
\newline
\indent While the model Hamiltonian captures the
band dispersion {for both the compounds} accurately, however the out of plane component of the ST obtained from DFT calculation have additional features which is not reproduced by $H_{L}^{C}$ suggesting the presence of symmetry
allowed additional terms.
\\
\indent Next we have analysed the top most valence bands around the L point for CoZrBi and SiLiIn that displays a maximum. In the absence of SOC the 3-fold degenerate $t_{2}$ bands at the $\Gamma$ point splits into a two fold degenerate and a singly degenerate band at the the L point; consistent with the $C_{3v}$ symmetry of the L point(see Fig. \ref{fig:figure2}(b) and \ref{fig:figure3}(b)). Inclusion of SOC further splits the four fold degenerate $J=\frac{3}{2}$ bands at the $\Gamma$ point into a pair of spin-orbit entangled doubly degenerate bands at the L-point (see Fig. \ref{fig:figure2}(d) and \ref{fig:figure3}(d)). From the Fig. \ref{fig:figure2}(b) and \ref{fig:figure3}(b) we find that the top of the valence band at the L point for CoZrBi is at the Fermi level, while for SiLiIn it lies about 1.7 eV below the VBM at the $\Gamma$ point. So in the following we have discussed the SOC induced band structure and consequent spin texture only for the 18-electron compound CoZrBi.
\begin{figure}[t]
\includegraphics[width=7.8cm,height=9.3cm,keepaspectratio]{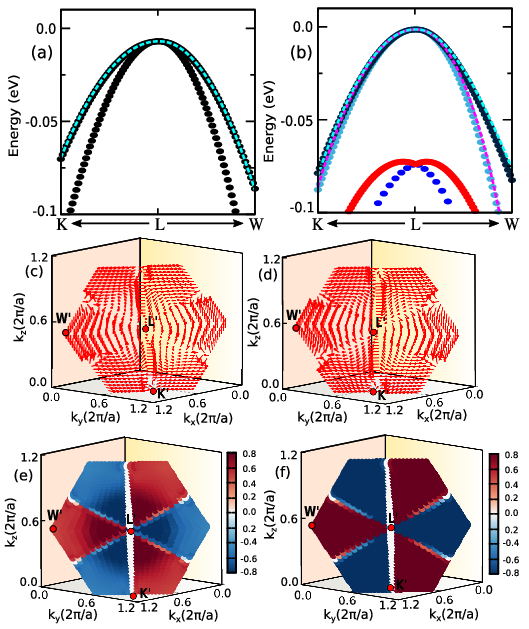}\caption{\label{fig:figure6} ((a), (b)) Band structure without and with SOC for the valence band maxima
for CoZrBi is plotted along $\frac{2\pi}{a}(0.538,0.538,0.425)$-$(\frac{2\pi}{a}(0.5,0.5,0.5)$-$\frac{2\pi}{a}(0.573,0.427,0.5)$
which lies along the path K$\leftarrow$L$\rightarrow$W. Band structure
obtained form DFT is plotted with dots and the band structure obtained
from the $\boldsymbol{k.p}$ model Hamiltonian is plotted with dashed
lines. ((c), (d)) ST of inner and outer branches around L$^{\prime}$
point for CoZrBi for the conduction band obtain from DFT calculation.
((e), (f)) Out-of-plane spin component of CoZrBi in the plane defined
by $k_{x}+k_{y}+k_{z}=\frac{3\pi}{a}$, obtained from DFT and model
Hamiltonian respectively.}
\end{figure}
\\ 
\indent In order to obtain further insights the DFT band dispersion of the top most valence bands are plotted in a narrow k-range around the neighborhood of the high symmetry L$\left(\frac{\pi}{a},\frac{\pi}{a},\frac{\pi}{a}\right)$ point along W$\left(\frac{2\pi}{a},0,\frac{\pi}{a}\right)$ and K$\left(\frac{3\pi}{2a},\frac{3\pi}{2a},0\right)$ direction without and including SOC as shown in Fig. \ref{fig:figure6}(a) and Fig. \ref{fig:figure6}(b) respectively. Fig. \ref{fig:figure6}(a) reveal a pair of doubly degenerate bands at the L point and the degeneracy is lifted upon inclusion of SOC (see Fig. \ref{fig:figure6}(b)). It is interesting to note from Fig. \ref{fig:figure6}(b) that the maxima of the topmost valence bands does not bifurcate away from the L point and remain at the L point, while for the lower two bands the maxima is bifurcated away from the L point displaying band crossing as expected for linear Rashba effect. This clearly establishes absence of linear Rashba effect for the tomost valence bands and therefore the leading term of the SOC Hamiltonian for the top most bands is expected to be cubic\citep{PhysRevLett.125.216405,LaWN3}.
\\
\indent The plot of the DFT spintextures for the two bands around the L$^\prime$ point is illustrated in Fig. \ref{fig:figure6}(c) and Fig. \ref{fig:figure6}(d). Our calculations reveal that the in-plane spin-component exhibit characteristic chiral configuration with the presence of substantial out-of-plane spin component as illustrated in Fig. \ref{fig:figure6}(e). In fact the out-of-plane component dominates over the in-plane component of $\langle{\vec{S}}\rangle$. 
\\
\indent In order to corroborate our DFT results a $\boldsymbol{k.p}$ model Hamiltonian is constructed with cubic Rashba as the leading term respecting the $C_{3v}$ symmetry at the L point \citep{LaWN3}. The symmetry adapted low energy $\boldsymbol{k.p}$ model Hamiltonian is : 
\begin{eqnarray}
H{^v_L}\left(k\right) &= & H_{0}\left({k}\right
)+\gamma\left(k^{{\prime}3}_x-3k^{{\prime}}_xk^{{\prime}2}_y\right)\sigma^\prime_z \nonumber \\
& & +\delta\left(\left(k^{{\prime}3}_x+k^{{\prime}}_xk^{{\prime}2}_y\right)\sigma^\prime_y-\left(k^{{\prime}3}_y+k^{{\prime}}_yk^{{\prime}2}_x\right)\sigma^\prime_x\right) 
\label{eq:equation3}
\end{eqnarray}
\\
\indent The band structure without and with SOC obtained from the model Hamiltonian for the topmost two bands are shown with dotted line in Fig. \ref{fig:figure6}(a) and \ref{fig:figure6}(b) respectively, and is in good agreement with DFT results. The calculated parameters of the model Hamiltonian are $\gamma$=5.5 $\mathrm{\mathrm{eV}}\mathring{A}^3$ and $\delta$=0.9 $\mathrm{\mathrm{eV}}\mathring{A}^3$ emphasizing the importance of the out-of-plane spin component. The out of plane component of the spin texture obtained from the model Hamiltonian shown in Fig. \ref{fig:figure6}(f) is in reasonable agreement with the DFT spin texture shown in Fig. \ref{fig:figure6}(e). The discovery of pure cubic Rashba splitting in non-polar half-Heusler alloys is rather unique and expected to find application in spintronics.
\\
\indent Similar results are obtained for the 8-electron compound SiLiIn where the splitting of the bands as expected is small.
\begin{figure*}
\includegraphics[width=18cm,height=8.5cm,keepaspectratio]{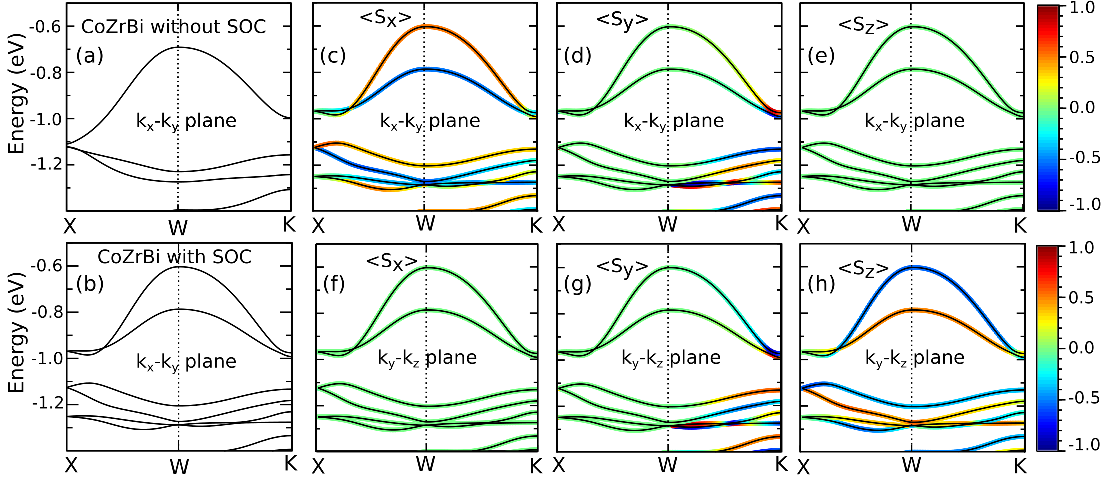}\caption{\label{fig:figure7}(a), (b)) Band structure without and with SOC in the valence band for CoZrBi is plotted along $\frac{2\pi}{a}(0.575,0.575,0.349)$-$(\frac{2\pi}{a}(0.5,0.5,0.5)$-$\frac{2\pi}{a}(0.648,0.351,0.5)$
which lies along the path K$\leftarrow$L$\rightarrow$W. Band structure
obtained form DFT is plotted with dashed
lines and the band structure obtained
from the $\boldsymbol{k.p}$ model Hamiltonian is plotted with dots. ((c), (d)) ST of inner and outer branches around L$^{\prime}$
point for CoZrBi for the conduction band obtain from DFT calculation.
((e), (f)) Out-of-plane spin component of CoZrBi in the plane defined
by $k_{x}+k_{y}+k_{z}=\frac{3\pi}{a}$, obtained from DFT and model
Hamiltonian respectively. ((g), (h)) Band structure without and with
SOC for SiLiIn is plotted along $\frac{2\pi}{a}(0.563,0.563,0.374)$-$(\frac{2\pi}{a}(0.5,0.5,0.5)$-$\frac{2\pi}{a}(0.623,0.377,0.5)$
which lies along the path K$\leftarrow$L$\rightarrow$W. ((i),(j))
ST of inner and outer branches around L$^{\prime}$ point for SiLiIn
in the conduction band. (k), (l) Out-of-plane spin component of SiLiIn
in the plane defined by $k_{x}+k_{y}+k_{z}=\frac{3\pi}{a}$, obtained
from DFT and model Hamiltonian respectively.}
\end{figure*}
\subsection{Zeeman spin splitting}
Till now we have discussed spin splitting and consequent spin textures
around the time reversal invariant k-points, here we shall consider
spin textures around the non time reversal invariant W point. The
band structure of the top of the valence band of CoZrBi around the
W$\left(\frac{\pi}{a},\frac{2\pi}{a},0\right)$ point along the path
X$\left(0,\frac{2\pi}{a},0\right)$ and K$\left(\frac{3\pi}{2a},\frac{3\pi}{2a},0\right)$
all lying in the $k_{x}$-$k_{y}$ plane without and including spin
orbit coupling are displayed in Figs. \ref{fig:figure7}(a) and \ref{fig:figure7}(b)
respectively. We find that the spin degeneracy of the bands are lifted
upon inclusion of SOC. In contrast to the Rashba and Dresselhaus effect
the spin splitting around the non-time reversal invariant W point
does not have band crossing rather the splitting is identical to that
realised in magnetic systems (i.e. in the absence of time reversal
symmetry). Interestingly such a splitting is now realised in a non-magnetic
system in the absence of inversion symmetry around a non-time reversal
invariant high symmetry W point and will be designated as Zeeman splitting.
It may be noted while the band structure around the W point in the
$k_{x}$-$k_{y}$ plane along the path X-W-K is identical to that
plotted along X$\left(0,\frac{2\pi}{a},0\right)$-W$\left(0,\frac{2\pi}{a},\frac{\pi}{a}\right)$-K$\left(0,\frac{3\pi}{2a},\frac{3\pi}{2a}\right)$
lying in the $k_{y}$-$k_{z}$ plane, the ST of the bands however
are dominated by different spin components depending on the chosen
plane. This is illustrated in Figs. \ref{fig:figure7}{[}(c)-(e){]}
for the $k_{x}$-$k_{y}$ plane and Fig. \ref{fig:figure7}{[}(f)-(h){]}
for $k_{y}$-$k_{z}$ plane where the primary contribution to the
ST is from $S_{x}$ and $S_{z}$ components respectively. \\
 \indent In order to explain the above observation we have calculated
the effective SOC term allowed by symmetry around the W point. The
point group symmetry around the W point is $S_{4}$, which contains
a two-fold rotation ($C_{2}$) around the principle axis. Additionally,
there are two four-fold rotations ($C_{4}$) around the principal
axis, one in the clockwise direction ($C_{4}^{+}$) and the other
in the anticlockwise direction ($C_{4}^{-}$), followed by a reflection.
The SOC term and the resulting SOC Hamiltonian \citep{MeraAcosta2019}
under the symmetry is, 
\begin{figure*}
\includegraphics[width=18cm,height=8.5cm,keepaspectratio]{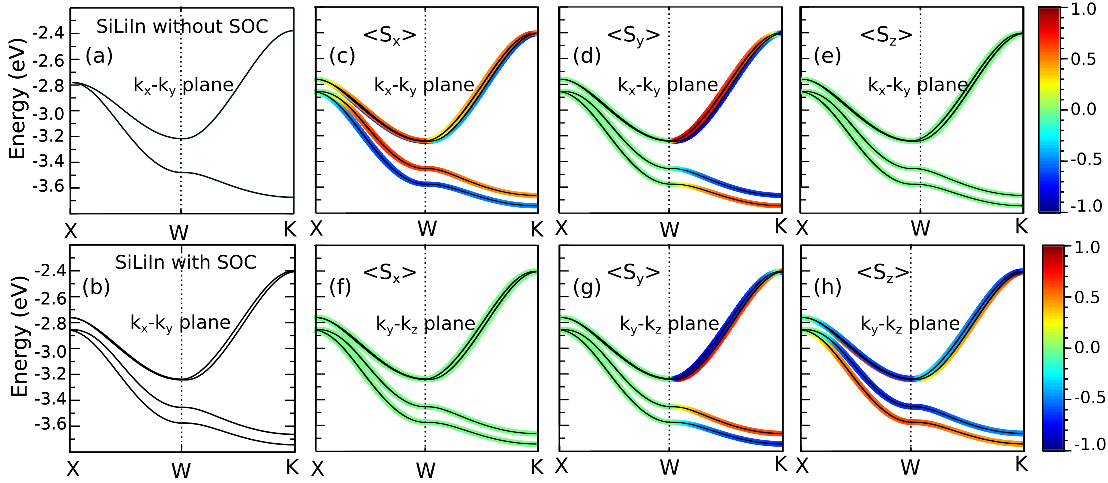}\caption{\label{fig:figure8} Figure shows the band structure along with projected
spin texture at non-time-reversal invariant $k$-point W of SiLiIn.
Panels (a) and (b) show the band structure of SiLiIn with and without
SOC along the path X$\left(0,\frac{2\pi}{a},0\right)$-W$\left(\frac{2\pi}{a},\frac{\pi}{a},0\right)$-K$\left(\frac{3\pi}{2a},\frac{3\pi}{2a},0\right)$.
Panels ((c)-(e)) show the band structure with projected spin textures
in the presence of SOC for SiLiIn along the path X$\left(0,\frac{2\pi}{a},0\right)$-W$\left(\frac{\pi}{a},\frac{2\pi}{a},0\right)$-K$\left(\frac{3\pi}{2a},\frac{3\pi}{2a},0\right)$
in the $k_{x}-k_{y}$ plane. Panels ((f)-(h)) show the band structure
and spin textures in presence of SOC for SiLiIn along the path X$(0,\frac{2\pi}{a},0)$-W$(0,\frac{2\pi}{a},\frac{\pi}{a})$-K$\left(0,\frac{3\pi}{2a},\frac{3\pi}{2a}\right)$
in the $k_{y}-k_{z}$ plane. The color code represents the orientation
of the spin component.}
\end{figure*}
\begin{eqnarray}
\Omega_{Z}\left(k\right) & = & \lambda_{Z}\left[k_{x}\left(k_{y}^{2}-k_{z}^{2}\right),k_{y}\left(k_{z}^{2}-k_{x}^{2}\right)\right.\nonumber \\
 &  & ,\left.k_{z}\left(k_{x}^{2}-k_{y}^{2}\right)\right]\nonumber \\
H_{W} & = & \mathbf{\Omega_{Z}\left(k\right)}.\boldsymbol{\sigma}\label{eq:equation4}
\end{eqnarray}
\\
 \indent where $\lambda{}_{Z}$ is the Zeeman parameter. At the W
point the splitting of the top two bands of CoZrBi is 186 meV as revealed
from our DFT calculations. As the W point lies at the boundary of
the BZ, therefore from the above Hamiltonian we can easily understand
that effective magnetic field is more at the boundary, causing a large
splitting.
\indent In order to understand the origin of SOC induced different
spin textures in different planes we need to understand the symmetry
associated with the chosen paths in the reciprocal space. The symmetry
operations around the W point in the $k_{x}$-$k_{y}$ plane along
X$\left(0,\frac{2\pi}{a},0\right)$-W$\left(\frac{\pi}{a},\frac{2\pi}{a},0\right)$
is $C_{2x}$. The spin components under the symmetry operation $C_{2x}$
transforms like, $C_{2x}:(\sigma_{x},\sigma_{y},\sigma_{z})\rightarrow(\sigma_{x},-\sigma_{y},-\sigma_{z})$,
ensuring primarily the $\sigma_{x}$ component survives in the Hamiltonian.
Using the model Hamiltonian in the $k_{x}$-$k_{y}$ plane with $k_{z}=0$,
we focus on the X$\left(0,\frac{2\pi}{a},0\right)$ $\rightarrow$
W$\left(\frac{\pi}{a},\frac{2\pi}{a},0\right)$ direction, where $k_{y}$
is fixed and only $k_{x}$ changes. This leads to an effective Hamiltonian
around the W point and along the path XW, given by $H_{W}=\lambda_{Z}k_{x}k_{y}^{2}\sigma_{x}$,
assuming $\left(k_{x}<<k_{y}=\frac{2\pi}{a}\right)$. This suggest
that the spin expectation value will be mostly contributed by the
$S_{x}$ component having positive value for the upper valence band
around the W point, and negative for the lower valence band in agreement
with DFT calculations shown in Figs. \ref{fig:figure7} {[}(c)-(e){]}.
Similarly for the the $k_{y}$-$k_{z}$ plane, the Hamiltonian with
$\left(k_{z}<<k_{y}=\frac{2\pi}{a}\right)$ will be given by $H_{W}=-\lambda_{Z}k_{z}k_{y}^{2}\sigma_{z}$,
as we move along the X$\left(0,\frac{2\pi}{a},0\right)$ $\rightarrow$
W$\left(0,\frac{2\pi}{a},\frac{\pi}{a}\right)$ direction where the
valence band primarily has $\sigma{}_{z}$ character. Here the expectation
value of $\langle S{}_{z}\rangle$ is negative for the top of the
valence band, while it is positive for the the lower valence band
in agreement with our DFT results.
\\
\indent The results of our calculation for the 8 electron half-Heusler
compound is shown in Fig. \ref{fig:figure8}. A pair of valence bands
around the W point exhibit Zeeman splitting upon application of SOC,
where the splitting is more for the lower pair of bands. It is to
be noted that splitting is weak in comparison to the 18 electron compound.
Further the nature of the spin textures are quite different from the
18 electron compound. In contrast to the 18 electron compound in the
$k_{x}$-$k_{y}$ plane, in addition to the $\langle S{}_{x}\rangle$
component there is appreciable $\langle S{}_{y}\rangle$ character
{[}see Fig. \ref{fig:figure8}(c-e){]}. Similarly in the $k_{y}$-$k_{z}$
plane in addition to the $\langle S{}_{z}\rangle$ character $\langle S{}_{y}\rangle$
character is also present{[}see Fig. \ref{fig:figure8}(f-h){]}.
\begin{figure*}[!t]
\includegraphics[width=16.5cm,height=8cm,keepaspectratio]{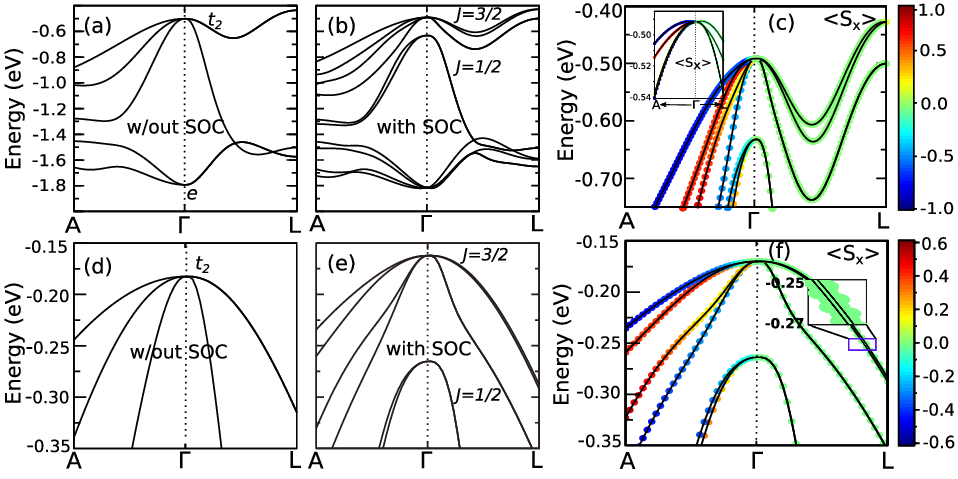}\caption{\label{fig:figure9} Band splitting with vanishing spin polarization
for 18- and 8 electron half-Heusler compounds. Panels (a) and (b)
show the band structure of CoZrBi with and without SOC along the path
A$\left(\frac{2\pi}{a}0.5,\frac{2\pi}{a}0.25,0.0\right)$-$\Gamma\left(0,0,0\right)$)-L$\left(\frac{2\pi}{a}0.5,\frac{2\pi}{a}0.5,\frac{2\pi}{a}0.5\right)$.
Panels (d) and (e) show the band structure of SiLiIn with and without
SOC along the path A$\left(\frac{2\pi}{a}0.5,\frac{2\pi}{a}0.25,0.0\right)$-$\Gamma\left(0,0,0\right)$)-L$\left(\frac{2\pi}{a}0.5,\frac{2\pi}{a}0.5,\frac{2\pi}{a}0.5\right)$.
Panels (c) and (f) show the band structure and the x component of
spin polarization for 18 electron and 8 electron compounds obtained
from DFT, inset of this (c) represents the band structure and the
x component of spin polarization obtain from $\boldsymbol{k.p}$ model
Hamiltonian.}
\end{figure*}
\subsection{Band splitting with vanishing spin polarization(BSVSP)}
The two half-Heusler systems considered here also exhibit another
intriguing phenomenon namely band splitting with vanishing spin polarisation,
where SOC splits the energy bands however both the split bands does
not exhibit net spin polarisation along certain high symmetry directions
of the BZ due to the presence of additional symmetries. In Fig. \ref{fig:figure9}(a)
the top of the valence band without SOC for CoZrBi along the path
A$\left(\frac{\pi}{a},\frac{\pi}{2a},0\right)$-$\Gamma\left(0,0,0\right)$-L
$\left(\frac{\pi}{a},\frac{\pi}{a},\frac{\pi}{a}\right)$ is displayed.
In the absence of SOC the $T_{d}$ symmetry at the $\Gamma$ point
splits the d bands into two fold degenerate $e$ and three fold degenerate
$t_{2}$ states. In Fig. \ref{fig:figure9}(b) the same band structure
including SOC is displayed. Inclusion of SOC further splits the $t_{2}$
states into two fold degenerate $J=\frac{1}{2}$ and four fold degenerate
$J=\frac{3}{2}$ states, with the $J=\frac{1}{2}$ states lying lower
in energy. Along the symmetry line $\Gamma$-L the $J=\frac{3}{2}$
states further split into three bands, where the lowest of the three
is doubly degenerate. In Fig. \ref{fig:figure9}(c) we have shown
the band structure projecting the $\langle S{}_{x}\rangle$ component
of the spin. The expectation value of $\langle S{}_{x}\rangle$ vanishes
for the top two spin split bands along $\Gamma$-L and the same is
true for the other spin components. The vanishing expectation value
of the spin $\langle\vec{S}\rangle$ for the two non-degenerate bands
along $\Gamma$-L over each Bloch wave function leads to BSVSP. \\
 \indent In order to understand the origin of the vanishing spin
polarisation we have constructed a $\boldsymbol{k.p}$ model Hamiltonian
following Ref. \citep{PhysRev.186.824,BSVSP}. The $\boldsymbol{k.p}$
Hamiltonian for the $J=\frac{3}{2}$ manifold of the top of the valence
band for the 18 electron compound CoZrBi in the vicinity of $\Gamma$
point may be written as $H=H^{+}+H^{-}$, where $H^{+}$ is invariant
with respect to the spatial inversion, 
\begin{eqnarray}
H^{+} & = & \frac{h^{2}}{m}\left[\left(\gamma_{1}+\frac{5}{2}\gamma_{2}\right)\frac{1}{2}k^{2}-\gamma_{2}\left(k_{x}^{2}J_{x}^{2}+k_{y}^{2}J_{y}^{2}+k_{z}^{2}J_{z}^{2}\right)\right]\nonumber \\
 &  & -2\gamma_{3}\left(\left\{ k_{x},k_{y}\right\} \left\{ J_{x},J_{y}\right\} +\left\{ k_{y},k_{z}\right\} \left\{ J_{y},J_{z}\right\} \right.\nonumber \\
 &  & +\left.\left\{ k_{z},k_{x}\right\} \left\{ J_{z},J_{x}\right\} \right)\label{eq:equation5}
\end{eqnarray}
and $H^{-}$ breaks the spatial inversion symmetry, 
\begin{eqnarray}
  H^{-} &  & =-\frac{2C}{\sqrt{3}}\left[k_{x}\left\{ J_{x},V_{x}\right\} +k_{y}\left\{ J_{y},V_{y}\right\} \right.\nonumber \\
 &  & \left.+k_{z}\left\{ J_{z},V_{z}\right\}\right]\label{eq:equation6}
\end{eqnarray}
Here, $\gamma_{1}$, $\gamma_{2}$, $\gamma_{3}$ and c are constants.
$J_{x}$, $J_{y}$ and $J_{z}$ are $4\times4$ angular momentum matrices
for a state of spin $\frac{3}{2}$, and $k_{x}$, $k_{y}$ and $k_{z}$
are the kinetic momentum terms. The symbol $\left\{ a,b\right\} $
means the symmetrized product $\frac{1}{2}\left(ab+ba\right)$. The
quantities $V_{x}$,$V_{y}$ and $V_{z}$ are given by: $V_{x}=J_{y}^{2}-J_{z}^{2}$,
$V_{y}=J_{z}^{2}-J_{x}^{2}$, $V_{z}=J_{x}^{2}-J_{y}^{2}$. Here,
we focus on the band degeneracy and spin polarization instead of the
exact band dispersion. The DFT band structure in a narrow range around
the $\Gamma$ point along the path A-$\Gamma$-L can be reasonably
reproduced with the fitted parameter values are $\gamma_{1}=-9.0$,
$\gamma_{2}=2.0$, $\gamma_{3}=-1.9$, $\frac{\hbar^{2}}{m}=1$ and
$C=-0.05$ {[}see Fig.\ref{fig:figure9}(c) inset{]}. The qualitative
results do not depend on the exact parameters. To investigate the
microscopic origin of the vanishing spin polarization along the symmetry
line $\Gamma$-L, we have calculated the on-site energy differences
between the orbitals of two non-degenerate $t_{2}$ states with opposite
spin orientations which can be regarded as the magnetic field acting
locally on the chosen orbitals. From the calculations we find that
the magnetic field on a given orbital vanishes. The local magnetic
moments associated with these orbitals cancel with each other, resulting
in a vanishing net spin polarization for the eigen state and vanishing
local spin polarization for the magnetic Co atom. In agreement with
the DFT results, the projection of the $\langle S{}_{x}\rangle$ component
on the band structure obtained from the model Hamiltonian also vanishes
supporting the validity of the low energy model Hamiltonian. \\
 \indent Similar results around the $\Gamma$ point and along the
path $\Gamma$-L is displayed by the 8 electron compound SiLiIn. In
Fig. \ref{fig:figure9}(d) the band structure without SOC is three-fold
degenerate around the $\Gamma$ point. Inclusion of SOC splits the
three-fold degenerate $t_{2}$ bands into $J=\frac{3}{2}$ quartet
and $J=\frac{1}{2}$ doublet as shown in Fig. \ref{fig:figure9}(e).
The splitting induced by SOC is 0.10 eV. Along the path $\Gamma$-L,
the top $J=\frac{3}{2}$ bands further splits into a pair of non-degenerate
bands and a doubly degenerate band and the former exhibit BSVSP effect
along $\Gamma$-L. The splitting is however much smaller in comparison
to the 18 electron compound. (see inset of Fig. \ref{fig:figure9}(f)).\\
 \indent Using the same model Hamiltonian as described for the 18
electron half-Heusler we can understand the origin of the BSVSP for SiLiIn. We
find the SOC-induced effective magnetic field acting on two different
p orbitals of the same atom are equal and opposite in strength along
the $\Gamma$L path, resulting in a BSVSP\citep{BSVSP}. { The splitting of the bands depends on the strength of SOC, as a consequence  BSVSP effect is much stronger (30 meV) for the 18-electron compound while it is 3 meV for the 8-electron compound SiLiIn and only 0.05 meV for GaAs, suggesting 18-electron compounds will be ideal candidate for experimental detection of BSVSP.}
\section{\label{sec:level5}CONCLUSION}
In the present paper we have analysed the electronic structure of
two representative half-Heusler systems with 18 electrons and 8 electrons
respectively in the presence of spin-orbit interaction. Our calculations
reveal rich features in the electronic structure due to spin-momentum
locking induced by SOC. Although both the compounds have identical
crystal structure, the orbital composition of the valence and conduction
bands are different for the 18 electron and the 8 electron system.
This brings in subtle changes in the SOC induced band structures emphasizing
the important role of orbitals. In addition, the BZ of the half-Heusler
system admits diverse local symmetries(little group) leading to non-trivial
splitting of the band structure due to SOC, resulting in novel spin
textures around the valleys of the high symmetry k-points. 
\begin{table}
\caption{\label{tab:table4} Our result and available Rashba and Dresselhaus
parameters in literature.}
\begin{ruledtabular}
\begin{tabular}{lcccc}
System & $\begin{array}{c}
\alpha_{R}\left(\mathrm{eV}\mathring{A}\right)\\
\mathrm{k-point}
\end{array}$ & \multicolumn{1}{c}{$\begin{array}{c}
\alpha_{D}\left(\mathrm{eV}\mathring{A}\right)\\
\mathrm{k-point}
\end{array}$} & $\begin{array}{c}
\text{\ensuremath{\mathrm{non}}-}\mathrm{polar}\\
\mathrm{or}\\
\mathrm{polar}
\end{array}$ & Reference\tabularnewline
\hline 
CoZrBi & $\begin{array}{c}
\mathrm{0.38}\\
\mathrm{L\text{\ensuremath{\left(\frac{\pi}{a},\frac{\pi}{a},\frac{\pi}{a}\right)}}}
\end{array}$ & $\begin{array}{c}
\mathrm{0.26}\\
\mathrm{X\text{\ensuremath{\left(0,0,\frac{2\pi}{a}\right)}}}
\end{array}$ & non-polar & This work\tabularnewline
SiLiIn & $\begin{array}{c}
\mathrm{0.15}\\
\mathrm{L\text{\ensuremath{\left(\frac{\pi}{a},\frac{\pi}{a},\frac{\pi}{a}\right)}}}
\end{array}$ & $\begin{array}{c}
0.08\\
\mathrm{X\text{\ensuremath{\left(0,0,\frac{2\pi}{a}\right)}}}
\end{array}$ & non-polar & This work\tabularnewline
BiTeI & $\begin{array}{c}
3.8\\
\Gamma\text{\ensuremath{\left(0,0,0\right)}}
\end{array}$ & $\begin{array}{c}
\mathrm{0.00}\\
\mathrm{\Gamma\text{\ensuremath{\left(0,0,0\right)}}}
\end{array}$ & polar & \citep{BiTeX_1}\tabularnewline
GeTe & $\begin{array}{c}
4.8\\
\mathrm{Z\left(\frac{\pi}{a},\frac{\pi}{a},\frac{\pi}{a}\right)}
\end{array}$ & $\begin{array}{c}
0.00\\
\mathrm{Z\left(\frac{\pi}{a},\frac{\pi}{a},\frac{\pi}{a}\right)}
\end{array}$ & polar & \citep{GeTe_1}\tabularnewline
HfO$_{2}$ & $\begin{array}{c}
0.028\\
\mathrm{T\left(0,\frac{\pi}{b},\frac{\pi}{c}\right)}
\end{array}$ & $\begin{array}{c}
0.578\\
\mathrm{\mathrm{T\left(0,\frac{\pi}{b},\frac{\pi}{c}\right)}}
\end{array}$ & polar & \citep{HfO2}\tabularnewline
LaWN$_{3}$ & $\begin{array}{c}
0.31\\
\mathrm{\Gamma\text{\ensuremath{\left(0,0,0\right)}}}
\end{array}$ & $\begin{array}{c}
0.01\\
\mathrm{\Gamma\text{\ensuremath{\left(0,0,0\right)}}}
\end{array}$ & polar & \citep{LaWN3}\tabularnewline
AlAs & $\begin{array}{c}
0.00\\
\mathrm{X\left(0,0,\frac{2\pi}{a}\right)}
\end{array}$ & $\begin{array}{c}
0.01\\
\mathrm{X\left(0,0,\frac{2\pi}{a}\right)}
\end{array}$ & non-polar & \citep{PhysRevB.72.195347}\tabularnewline
GaP & $\begin{array}{c}
0.00\\
\mathrm{X\left(0,0,\frac{2\pi}{a}\right)}
\end{array}$ & $\begin{array}{c}
0.07\\
\mathrm{X\left(0,0,\frac{2\pi}{a}\right)}
\end{array}$ & non-polar & \citep{PhysRevB.72.195347}\tabularnewline
\end{tabular}
\end{ruledtabular}
\end{table}
\\
\indent Around the X point we have observed Dresselhaus effect as
expected for a non centrosymmetric and non-polar material \citep{PhysRevB.72.195347}
{[}see Table \ref{tab:table4}{]}. Using a symmetry adapted $\boldsymbol{k.p}$
model Hamiltonian around the high symmetry X point, we have calculated
the band dispersion and the consequent ST. The model Hamiltonian only
required linear terms to reproduce the DFT results, and the Dresselhaus
parameter $\alpha_{D}$ for the 18 electron and the 8 electron system
is calculated to be 0.26 eV$\mathring{A}$ and 0.08 eV$\mathring{A}$
respectively, in good agreement with DFT results. \\
 \indent Similar to the Dresselhaus effect, we have observed the
Rashba effect in both half-Heusler systems around the polar L$\left(\frac{\pi}{a},\frac{\pi}{a},\frac{\pi}{a}\right)$
point in the {{[}111{]}} plane. The Rashba effect features both
linear and higher-order terms for the conduction band, but the strength of the higher-order
term is found to be much weaker compared to the linear term. { However for the topmost valence band at the L point we have identified  cubic Rashba  to be the leading term of the SOC Hamiltonian with novel spin texture. The identification of pure cubic Rashba splitting in half-Heusler alloys is rather unique and are expected to play important role in spin transport.} Further
both the Dresselhaus and Rashba splitting is found to be much weaker
for the 8 electron sp based compound in comparison to the 18 electron
system. \\
 \indent At the non-TR invariant high symmetry point W, we observe
Zeeman spin splitting. From the $\boldsymbol{k.p}$ model Hamiltonian,
we have established that only the cubic terms are invariant under
symmetry operations, and the component of spin orientation depends
on the chosen plane. \\
 \indent Finally, we have observed the BSVSP along the high symmetry
path $\Gamma$-L in both 18 electron and 8 electron half-Heusler systems.
Around the $\Gamma$ point the tetrahedral environment causes the
d orbitals of the 18 electron half-Heusler to split into $t_{2}$
and $e$ states while leaving the p states degenerate for the 8 electron
half-Heusler compound. The SOC further splits the $t_{2}$ and p states
into four fold degenerate $J=\frac{3}{2}$ and two fold degenerate
$J=\frac{1}{2}$ states. Along the path $\Gamma$-L the four fold
degenerate $J=\frac{3}{2}$ states further splits into two non degenerate
and a two fold degenerate bands. The presence of a pair of non degenerate
bands in the presence of SOC with the little group of the relevant
k-point belonging to the non-pseudo polar point group leads to BSVSP\citep{BSVSP}. 
\\
\indent { Our calculations reveal that the valence band and the conduction band of half-Heusler alloys at the various high symmetry points feature extrema(valleys) characterized by spin texture which are dependent on the location of the valleys in the k-space designated by a valley index. In the presence of SOC the spin splitting is tied to the valley index, requiring the scattering of charge carriers between valleys to have simultaneous spin flip and momentum transfer. This favors long valley life time required for valleytronics application. These valleys can be accessible by doping.} 
\\
 \indent Our detailed first principle calculations complemented with
$\boldsymbol{k.p}$ model Hamiltonian method for the two representative
half-Heusler systems identifies in a family of ternary half-Heusler system
with heavy elements { another novel functionality  for potential application}
in spin-valleytronics \textbf{\citep{Casper_2012}}.
\begin{acknowledgments}
K.D thanks Council of Scientific and Industrial Research(CSIR) for support through a fellowship(File No. 09/080(1178)/2020-EMR-I).
I.D. thanks Science and Engineering Research Board (SERB) India (Project
No. CRG/2021/003024) and Technical Research Center, Department of
Science and Technology(TRC-DST) for support. 
\end{acknowledgments}

%

\end{document}